# Classifications of Skull Fractures using CT Scan Images via CNN with Lazy Learning Approach

[1]Md Moniruzzaman Emon, [2]Tareque Rahman Ornob and [3]Moqsadur Rahman

[1]*Department of Computer Science and Engineering, Sylhet Engineering College, School of Applied Sciences and Technology, Shahjalal University of Science and Technology, Sylhet, Bangladesh*
[2]*Department of Computer Science and Engineering, School of Applied Sciences and Technology, Shahjalal University of Science and Technology, Sylhet, Bangladesh*
[3]*Department of Computer Science and Engineering, Shahjalal University of Science and Technology, Sylhet, Bangladesh*

**Abstract:** Classification of skull fracture is a challenging task for both radiologists and researchers. Skull fractures result in broken pieces of bone, which can cut into the brain and cause bleeding and other injury types. So it is vital to detect and classify the fracture very early. In real world, often fractures occur at multiple sites. This makes it harder to detect the fracture type where many fracture types might summarize a skull fracture. Unfortunately, manual detection of skull fracture and the classification process is time-consuming, threatening a patient's life. Because of the emergence of deep learning, this process could be automated. Convolutional Neural Networks (CNNs) are the most widely used deep learning models for image categorization because they deliver high accuracy and outstanding outcomes compared to other models. We propose a new model called SkullNetV1 comprising a novel CNN by taking advantage of CNN for feature extraction and lazy learning approach which acts as a classifier for classification of skull fractures from brain CT images to classify five fracture types. Our suggested model achieved a subset accuracy of 88%, an F1 score of 93%, the Area Under the Curve (AUC) of 0.89 to 0.98, a Hamming score of 92% and a Hamming loss of 0.04 for this seven-class multi-labeled classification.

## Introduction

In general, CT images and radiology reports provide additional information to a physician, making an informed decision. The radiologist studies the image and notes the findings in the traditional diagnosis procedure and the physician then choose a treatment based on the diagnosis. All this requires a lot of time. As a severe impact or blow to the head may cause a skull fracture, as well as brain injury, it is vital to quickly identify what lesion the skull fracture may cause to the brain to give the patient proper treatment. If the fracture occurs over a major blood vessel, significant bleeding may occur inside the brain, so head injury patients with skull fracture have far more intracranial hematomas than those without fractures (Zaki *et al*., 2008, Liu *et al*., 2008). Classification is a way to find out the resultant lesion from skull fractures. For suspected skull or brain injuries, Computed Tomography (CT) has become the standard diagnostic method for Acute Care [UK and others]. It is a medical imaging technique that incorporates advanced x-ray and computer technology. Brain CT scans provides more accurate information about brain tissue and structures than regular head X-rays, allowing for more information about brain injuries and diseases. It is also helpful for people who cannot get an MRI if their brains contain metal. The CT scans, as opposed to MRI, have a broader availability, are easier to use and can analyze unique structures of the brain to look for a mass, stroke, bleeding region, or blood vessel abnormality. Based on a CT scan, a radiologist determines whether there is any skull fracture and to which category the fracture belongs. However, the skull fracture has the following characteristics on CT scans: Fractures commonly appear as narrow slits, fractures might be found in various locations and lengths and a significant number of fractures are tiny.

All of those characteristics can make manual detection of skull fracture and its classification both time-consuming and difficult. As a result, presenting an effective automated skull fracture classification and detection system is critical. The automatic detection and the classification of skull fractures could aid in the detection of other abnormalities in the CT scan brain images. Also, many of the hospitals in the world are understaffed, which causes delays in evaluating CT scan images. In the CQ500 dataset, their three radiologists disagreed on many cases if a patient's skull is fractured or normal, let alone deterministically classify if that fracture is a calvarial fracture or another fracture. Automatic classification could help doctors in an understaffed hospital determine more critical patients to give them treatment first. There is only one approach as far as our knowledge of the automated detection of linear skull fracture. No one has automated the classification of common skull fractures. These facts provide the motivation for this study. Because of the advances of deep learning, several improvements in the use of deep learning for medical imaging interpretation tasks have been made in the past years. We present a novel model, SkullNetV1, for automated detection of common five skull fractures (i.e., linear, depressed, linear non-depressed, facial, comminuted) besides detecting fractured or normal cases from CT scan images. These benefits may form a solid foundation for content-based medical image retrieval for medical training or diagnosis.

## Related Works

Some methods for detecting skull fractures have already been introduced. Shao and Zhao (2003) have concentrated their efforts on CT brain segmentation for automated diagnosis of skull fractures. They have proposed a method for segmenting the brain image using a region-growing method and then used the entropy feature to generate rules for diagnosing skull fractures. This approach has shown a high rate of acceptance. However, its complexity and performance can both be simplified and improved computationally. Zaki *et al.* (2009) have applied the Sobel edge detection approach for the diagnosis of skull fractures. Despite the fact that the Sobel edge detection's approach is better for a variety of characteristics, it does produce some misshaping lines in some circumstances. It's worth noting that this approach can't handle huge features. The detection of the edges Prewitt is a simple method that is based on the gradient magnitude. However, as the gradient magnitude drops, the accuracy almost surely diminishes. Abubacker *et al.* (2013) provided a simple and fast automatic approach in Digital Imaging and Communications in Medicine (DICOM) to extract the skull bone and diagnose the fracture utilizing histogram-based thresholding and neighboring pixel connection search. This approach's experimental results are consistent, with a high detection rate. Chilamkurthy *et al.* (2018) have created a deep learning method for automated detection of intracranial hemorrhage and its types (i.e., intraparenchymal, intraventricular, subdural, extradural and subarachnoid); calvarial fractures, midline shift and mass effect. They showed that deep learning algorithms could perform this task with high accuracy. Kuang *et al.* (2020) have proposed a method to detect skull fractures more accurately in a short time. The proposed method is called Skull R-CNN. Compared to previous research on skull fracture detection, Skull R-CNN has fewer false positives while maintaining high sensitivity. Lee *et al.* (2020) presented an algorithm for detecting Femur fracture from pelvic X-ray images that uses available paired image-text training data (meta-training set) to learn features from both modalities without an additional hallucination network. When using this approach, a novel deep learning model is produced to operate only over the single-image modality input and outperforms the standard network trained only on image data. Thus, the new method transfers information commonly extracted from text training data to a network that can extract associated information from image counterparts. The key steps in the proposed approach include image normalization, centroid identification, multi-level global segmentation and, skull skeletonization. Yamada *et al.* (2016) developed a novel method to automatically detect linear skull fractures by detecting crack lines on head CT images. They performed a basic evaluation using two kinds of phantoms. Their experiment with a digital phantom revealed that a crack line with a width of 0.35 mM could be detected. However, all the approaches listed above only looked at the local features of the skull fractures. No one has classified the common types of skull fracture. Using head CT scans, we developed a novel deep learning technique to confidently detect and classify common skull fractures. In contrast to previous research that focused just on detecting skull fractures, our objective is to classify skull fractures from images. We introduce SkullNetV1, a new model for automated classification of the five most common skull fractures (linear, depressed, linear, non-depressed, facial and comminuted).

## Dataset

We collected head CT scan images of 232 patients from Medinova Medical Services Ltd. and Ibn Sina Hospital Sylhet Limited with their respective permission. Every CT scan image was in DICOM format, with 512×512 pixels in 0.75 mm, 1.0 mM and 5.0 mM slice thickness. CT scan images were anonymized automatically. All the collected 232 CT scans were examined and annotated by the top radiologists. First, the radiologists annotated the dataset into the fracture and the normal cases. Fractured cases were further classified into five classes by the respective radiologists. The paired data (174 CT scan images and radiology reports) were used as training sets.

## Methodology

DICOM image processing requires more computing power than JPG image processing, which we didn't have,

so all DICOM images were converted to JPG format. The image size was reduced to 200x200 pixels to save memory and speed up the training process. Pixel data was rescaled by dividing with maximum pixel value.

We consider this study as a multi-label classification problem to determine what types of fracture exist in a CT scan image, in addition to whether the CT scan image is fractured or not. Thus, each patient's CT scan picture slices were multi-labeled with zero or one in a CSV file. A NumPy array was created from the CSV file to encode the label data.

The goal of multi-label image classification is to extract all of an image's semantic classifications. Given an image I, the final prediction $l_k$ of the k-th class corresponding to I is formulated by

$$l_k = \text{I}(p_k(\text{I} \mid \text{w}) > \tau_k), k \in \{1,\cdots,K\} \quad (1)$$

where $p_k(I|w)$ signifies the posterior probability of image *I*, including the kth class, as estimated by a model with parameters *w*. The number of labels given is *K* and the confidence threshold for the kth class is $\tau_k$. $I(p > \tau)$ is an indicator function that returns 1 when p is greater than zero and 0 otherwise. The last label indication is $l_k$, which implies that if $l_k = 1$, then the kth class is present in the image and if $l_k = 0$, it is not.

The major goal of this phase is to extract characteristics from a CT scan image that will be sent into a classifier, which will categorize the CT image into one or more classes. A baseline model was created as the feature extractor that consisted of 8 convolutional layers with modest 3x3 filters followed by a max-pooling layer, with the number of filters doubling with each consecutive block. Each block has two convolutional layers with three filters, as well as Leaky ReLU activation at 0.001 learning rate and He weight initialization with the same padding, guaranteeing that the output feature maps are the same width and height. After that, a max-pooling layer with a 3x3 kernel was added. There are four of these blocks, each containing 32, 64, 128 and 256 filters. The final pooling layer's output was flattened to a one-dimensional vector to compute the Euclidean distance. The architecture, layer and number of parameters of our baseline CNN are given in Table 1.

Binary Relevance, Classifier Chains and Label Power set are three notable approaches to convert multi-label problems into single-label problems. Binary relevance treats each label as a separate single class classification; however, because each target variable is treated independently, it ignores label correlation. The first classifier in a Classifier Chain is trained just on the input data and then each subsequent classifier is trained on the input space as well as all prior classifiers in the chain; this overcomes the issue Binary Relevance has. Classifier Chains, on the other hand, will perform poorly if there is no label correlation. Label Power set transforms the problem into a multi-class problem by training a single multi-class classifier on all unique label combinations discovered in the training data. The Label Power set approach is arguably the finest of the three; the one problem is that as the training data grows, the number of classes grows as well. As a result, the model's complexity grows, which would result in lower accuracy.

We did not use any of these classifiers as our intention is to compare the Euclidean distance of the test image with all the images in the training set without transforming the problem into different subsets of problems. Then we check the image for which we got the minimum Euclidean distance and then the classifier predicts the most frequent labels.

ML-KNN Szymański and Kajdanowicz´ (2017) suits our purpose, which is an upgraded version of KNN. To improve performance, it uses the maximum a posteriori principle to label a new instance after finding the k nearest neighbors in the training data. Since ML-KNN operates on sparse matrices internally using SciPy sparse matrix library, it is highly memory-efficient.

In order to train the ML-KNN, the extracted features from CNN and label data were used as input. ML-KNN is known as a lazy learning algorithm because it does not learn anything in the training period. The lazy learning approach involved determining the most similar samples to each new query to be predicted from the entire training data set. Once a selection of training patterns has been chosen, the classification model is learned with that subset and used to categorize the new query. The value of k in ML-KNN, which is the number of neighbors of each input instance to consider, was set to 3. It's a work with an imbalanced number of labels in a multi-label classification system. Because there are so many classes to predict, the notion of positive and negative words and associated terms are calculated for each kind in a one vs. rest way and then the overall levels are averaged. As a result, we chose F1 scores to evaluate all models.

TP = True Positive, TN = True Negative, FP = False positive, FN = False Negative

$$accuracy(y,y) = \frac{1}{n_{samples}} \sum_{i=0}^{nSamples-1} 1(\hat{y}_i = y_i)$$

$$\text{Pr}ecision = \frac{TP}{(TP + FP)} \quad (2)$$

$$\text{Re}call = \frac{TP}{(TP + FN)} \quad (3)$$

$$F1 = 2 * \frac{\text{Pr}ecision * \text{Re}call}{\text{Pr}ecision + \text{Re}call} \quad (4)$$

Accuracy, Precision, Recall and AUC score were also added besides the F1 score.

## Experimental Environments

Ubuntu 20.04.2 LTS was utilized as the operating system. The AMD Ryzen Thread ripper 1950X 16-Core Processor was used. The primary memory was 32 GB and the graphics cards were two GeForce RTX 2080 *Ti* with 11 GB RAM each. TensorFlow 2.5.0rc2 was used as the deep learning framework.

## Experimental Results

The dataset was randomized and then divided into three parts: 60% Training, 20% validation and 20% testing.

Four transfer learning-based CNN models, an original Alex Net model and our baseline CNN was trained to check their performance.

### Binary Cross-Entropy Loss of Every CNN Model

From the Fig. 4, 5, 6, 7, 8, 9; it is clear that apart from our baseline CNN, every other CNN model shows major overfitting behavior. Table 2 further clarifies our argument. Table 2 shows a comparison of the most well-known CNN models and our suggested model's classification performance in terms of AUC and micro average F1 score. From Table 2 it is clear that our proposed model, SkullNetV1 beats other models by a large margin.

Average Precision graph is shown in Fig. 10 to show the area under the precision-recall curve from the proposed model. The weighted mean of precision reached at each threshold, with the increase in recall from the preceding threshold used as the weight, describes a precision-recall curve as Average Precision.

Precision-Recall curve of all classes computed by our model is shown in Fig. 11 to evaluate our proposed model's output quality.

To further check our proposed model's performance per class wise, Table 3 presents a summary of classification performance results in terms of precision, recall, F1 score, ROC AUC score and Specificity. It also includes our proposed model's score over all classes across the testing dataset.

Even though the dataset has an uneven label distribution (Fig. 1), our strategy outperforms state-of-the-art CNN models on the dataset, where SkullNetV1 learns substantially better than InceptionResNetV2 (InceptionResNetV2 is the next best model according to Micro Average F1 score); SkullNetV1 achieved an overall accuracy of 0.88 and a Micro average F1 score of 0.93 in this seven-class multi-label task. The SkullNetV1 showed good performance for all the performance metrics. Not only that, but our SkullNetV1 model outperformed the competition across the board.

**Table 1:** Architecture, layer, parameters of our baseline CNN

| SL No. | Layer | Activation shape | Activation size | Parameters |
|---|---|---|---|---|
| 1. | Input Layer | (200,200,3) | 120000 | 0 |
| 2. | CONV2d_1(filter_shape = 3, stride = 1, num_filters = 32) | (200,200,32) | 1280000 | 896 |
| 3. | CONV2d_2(filter_shape = 3, stride = 1, num_filters = 32) | (200,200,32) | 1280000 | 9248 |
| 4. | MaxPooling2d_1 | (100,100,32) | 320000 | 0 |
| 5. | CONV2d_3(filter_shape = 3, stride = 1, num_filters = 64) | (100,100,64) | 640000 | 18496 |
| 6. | CONV2d_4(filter_shape = 3, stride = 1, num_filters = 64) | (100,100,64) | 640000 | 36928 |
| 7. | MaxPooling2d_2 | (50,50,64) | 160000 | 0 |
| 8. | CONV2d_5(filter_shape = 3, stride = 1, num_filters = 128) | (50,50,128) | 320000 | 73856 |
| 9 | CONV2d_6(filter_shape = 3, stride = 1, num_filters = 128) | (50,50,128) | 320000 | 147584 |
| 10. | MaxPooling2d_3 | (25, 25, 128) | 80000 | |
| 11. | CONV2d_7(filter_shape = 3, stride = 1, num_filters = 256) | (25, 25, 256) | 160000 | 295168 |
| 12. | CONV2d_8(filter_shape = 3, stride = 1, num_filters = 256) | (25, 25, 256) | 160000 | 590080 |
| 13. | MaxPooling2d_4 | (12, 12, 256) | 36864 | 0 |
| 14. | Flatten | (36864) | | 0 |
| | Total params | 1172256 | | |

**Table 2:** Results from multiple models on the testing dataset

| Model | Measure | |
|---|---|---|
| | AUC | F1 score |
| InceptionV3 | 0.83 | 0.63 |
| Alex Net | 0.77 | 0.6 |
| ResNet50 | 0.82 | 0.62 |
| Efficient Net | 0.84 | 0.62 |
| InceptionResNetV2 | 0.85 | 0.65 |
| Baseline CNN (ours) | 0.9 | 0.75 |
| SkullNetV1 (proposed model) | 0.94 | 0.93 |

**Table 3:** Class wise score of the proposed model

|  | Labels | Precision | Recall | F1 score | ROC AUC | Specificity |
|---|---|---|---|---|---|---|
|  | Fracture | 0.98 | 0.98 | 0.98 | 0.98 | 0.92 |
|  | Not fractured | 0.97 | 0.97 | 0.97 | 0.98 | 0.98 |
|  | Linear fracture | 0.97 | 0.83 | 0.90 | 0.91 | 0.99 |
|  | Depressed fracture | 0.91 | 0.92 | 0.91 | 0.94 | 0.95 |
|  | Linear non-depressed | 0.92 | 0.79 | 0.85 | 0.89 | 0.98 |
|  | Fracture facial fracture | 0.97 | 0.82 | 0.89 | 0.91 | 0.99 |
|  | Comminuted Fracture | 0.93 | 0.85 | 0.89 | 0.92 | 0.98 |
| Micro avg |  | 0.96 | 0.91 | 0.93 | 0.94 | 0.96 |
| Macro avg |  | 0.95 | 0.88 | 0.91 | 0.93 | 0.94 |
| Weighted avg |  | 0.96 | 0.91 | 0.93 | 0.94 | 0.93 |
| Samples avg |  | 0.96 | 0.93 | 0.94 | 0.95 |  |
|  | Score of SkullNetV1 over all classes |  |  |  |  |  |
| Subset Accuracy |  | 0.88 |  |  |  |  |
| F1 score |  | 0.93 |  |  |  |  |
| Hamming Score |  | 0.92 |  |  |  |  |
| Hamming Loss |  | 0.04 |  |  |  |  |
| ROC AUC |  | 0.94 |  |  |  |  |

**Table 4:** Comparative performance of published model vs. proposed model

| Paper | Objective | Score, metric | Dataset |
|---|---|---|---|
| Shao and Zhao (2003) | Automatically detect if the skull is fractured or not | 100%, accuracy | Training dataset = 100 images, testing dataset = 100 images |
| Zaki *et al*. (2009) | Segment fractured skull from 2D-CT brain image | 95%, Normalized Recall rate | 3004 normal and 28 original fracture cases |
| Yamada *et al*. (2016) | Detection of Linear skull fracture | 80% accuracy for a crack line of width 1.05 mm | 3D CT scan |
| Chilamkurthy *et al*. (2018) | Detection of multiple Hemorrhage and skull fracture of only calvarial. | 91.11%, AUC | 313318 slices of head CT scans |
| Lee *et al*. (2020) | Detection of femur Fracture | 86.78%, accuracy | Training dataset = 459 cases, testing dataset = 227 cases |
| Kuang *et al*. (2020) | Faster detection of skull fracture more accurately | 80%, precision recall score | 45 CT scan comprising of 872 slices |
| Ours | Classification of five common non-exclusive skull fractures | 93%, F1-score | Training dataset = 174 cases (24153 CT scan slices), testing dataset = 54 cases (8051 CT scan slices) |

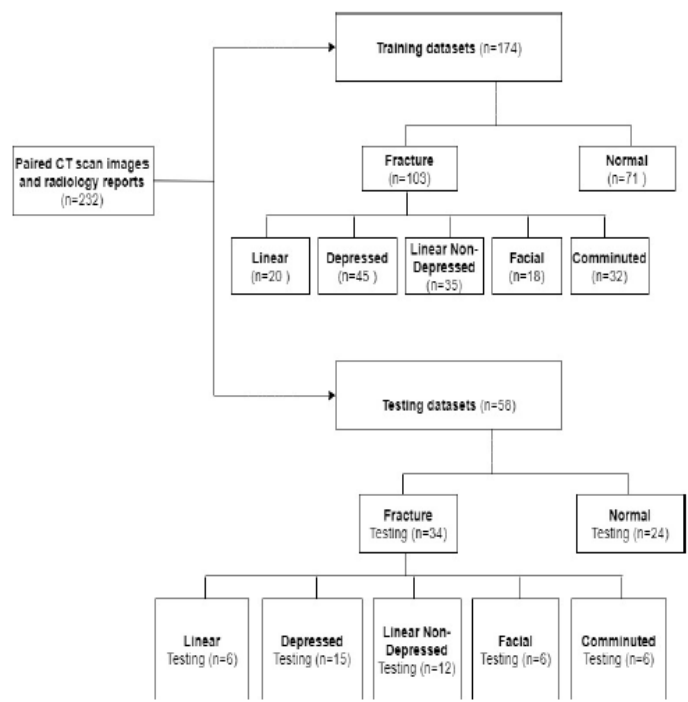

**Fig. 1:** Data characteristics. Number of CT images and radiology reports for training and testing the system.

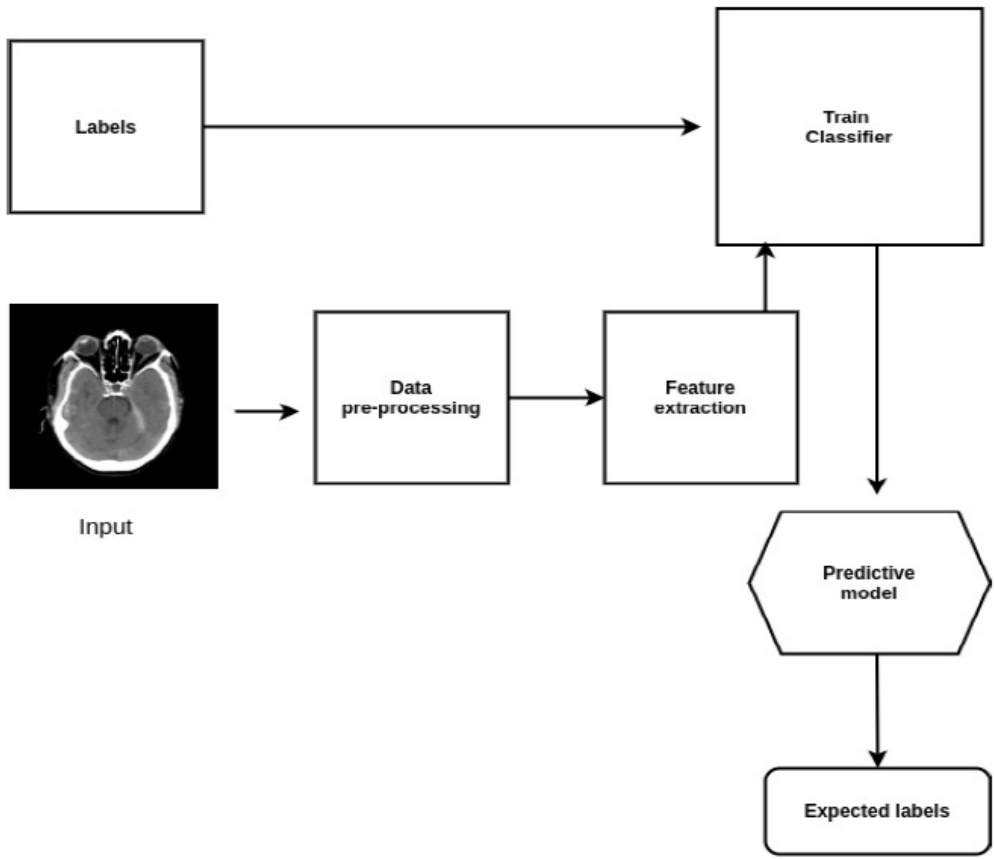

**Fig. 2:** Multi labeled skull fracture classification pipeline

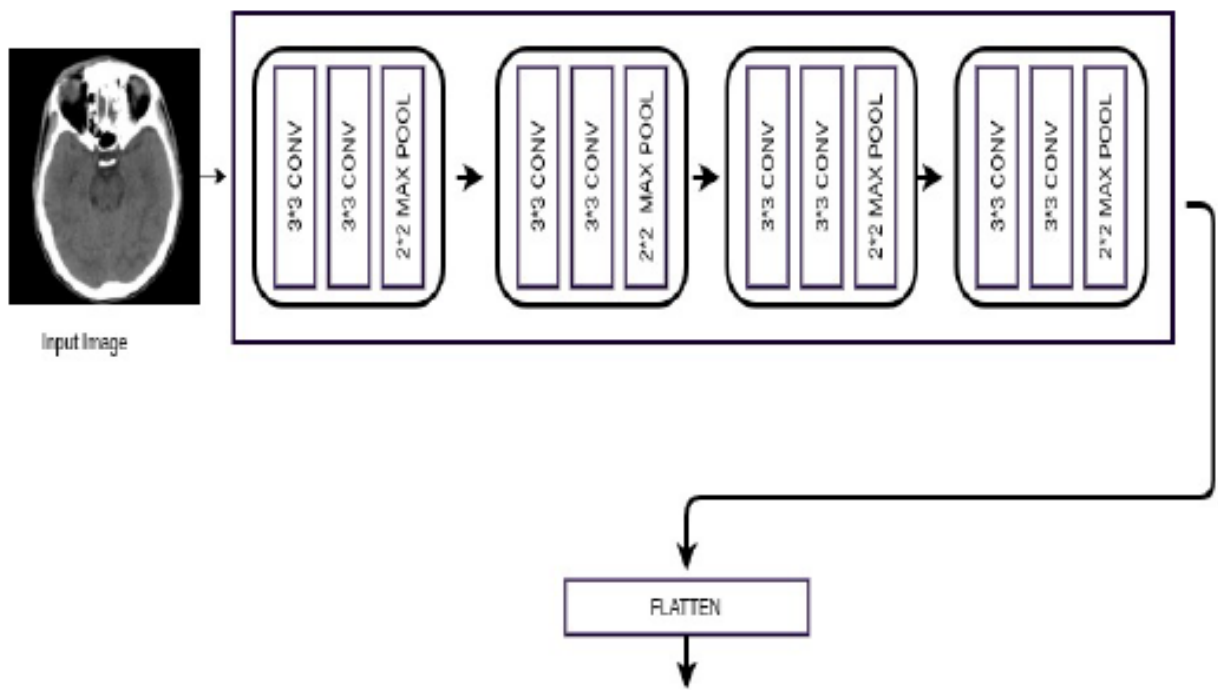

**Fig. 3:** Proposed architecture of feature extractor

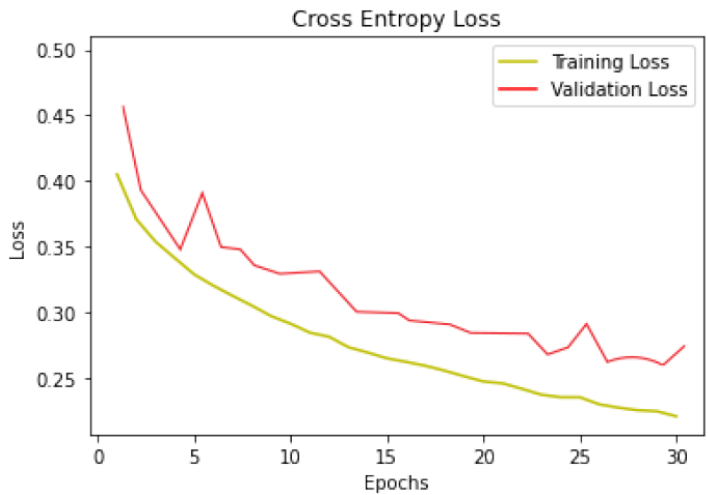

**Fig. 4:** InceptionV3

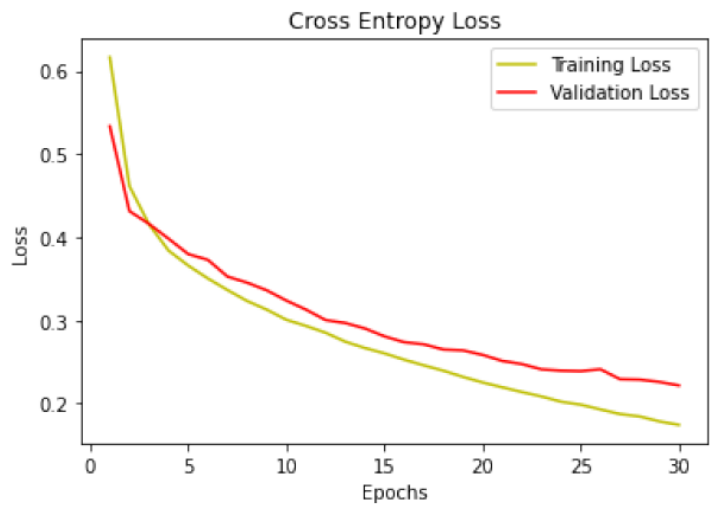

**Fig. 5:** AlexNet

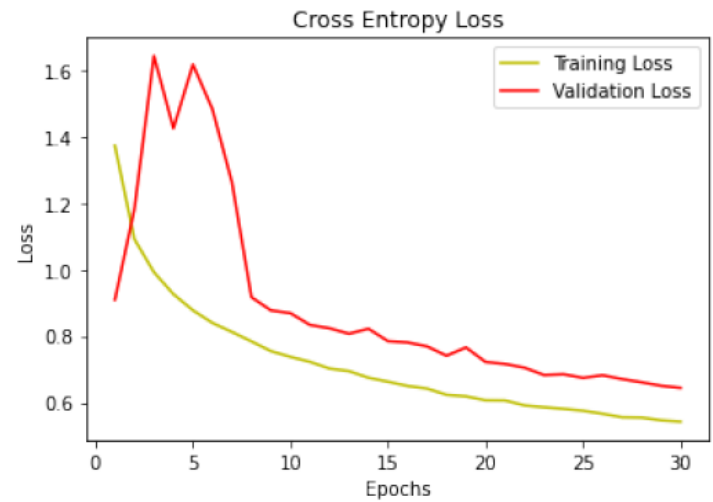

**Fig. 6:** ResNet50

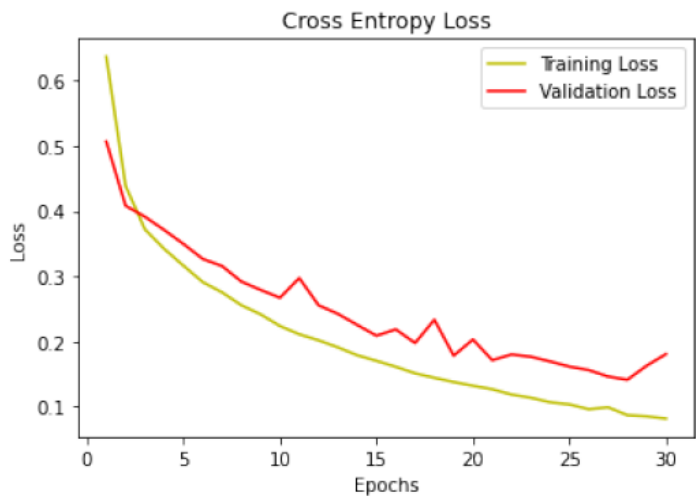

**Fig. 7:** EfficientNet

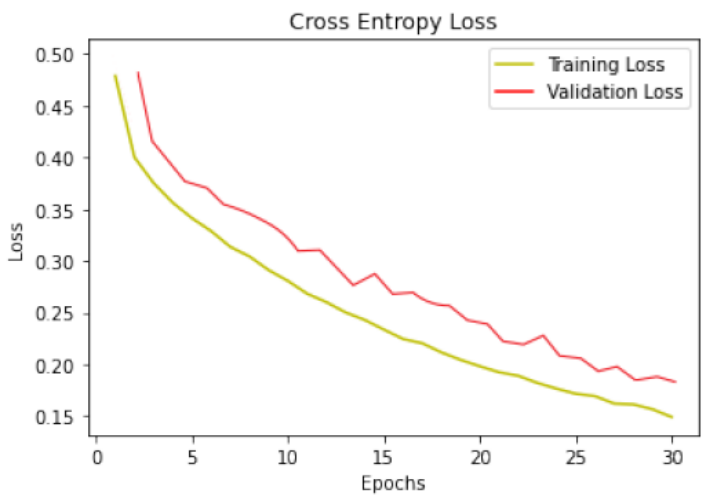

**Fig. 8:** InceptionResNetV2

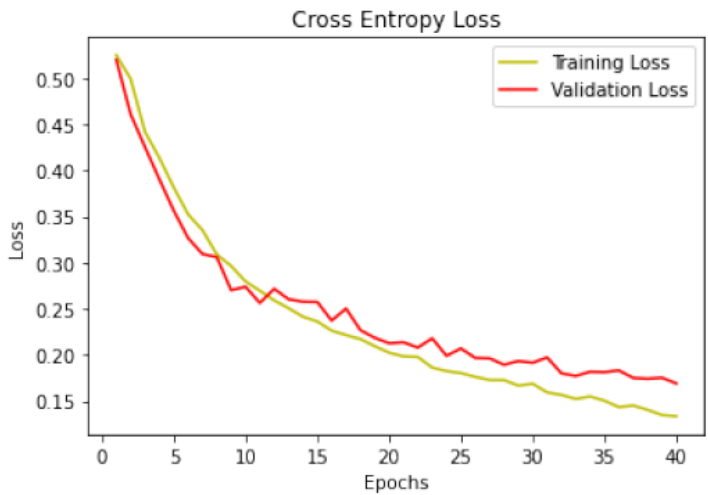

**Fig. 9:** Baseline CNN (ours)

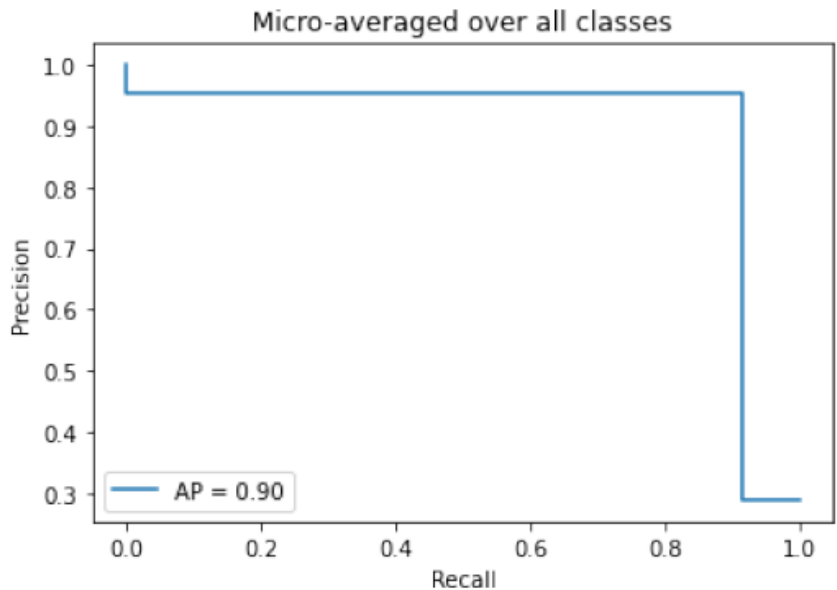

**Fig. 10:** Average Precision graph over all classes

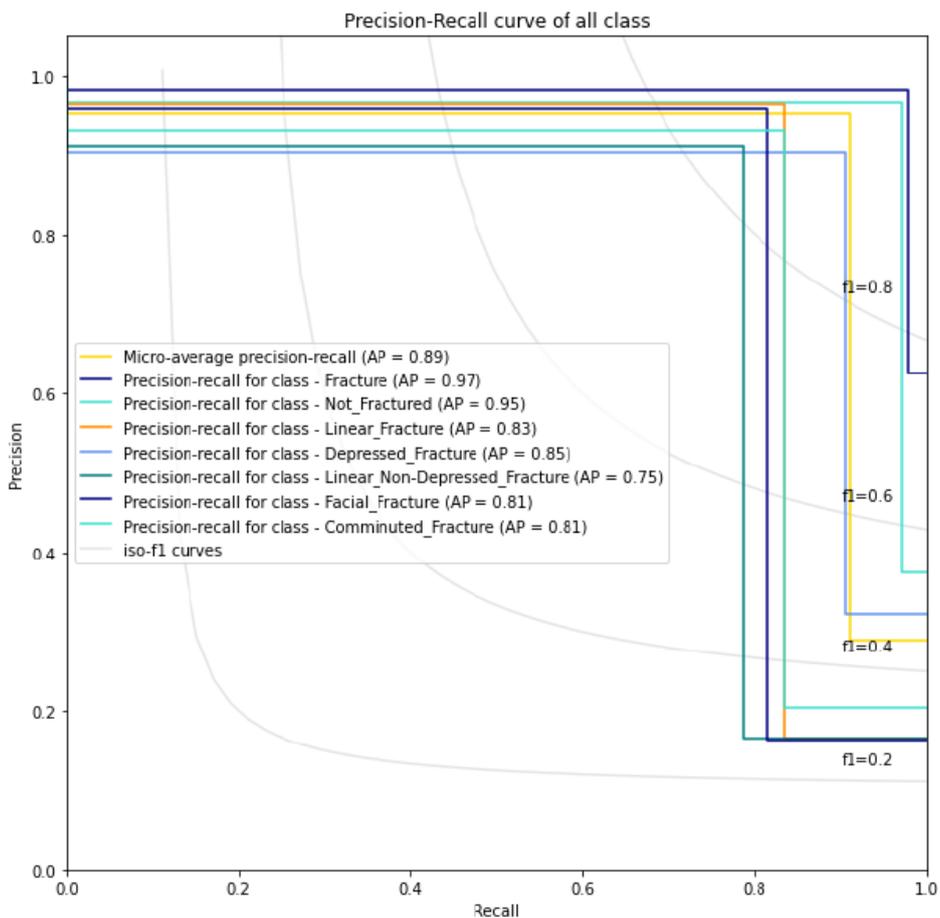

**Fig. 11:** Precision-recall curve of all classes

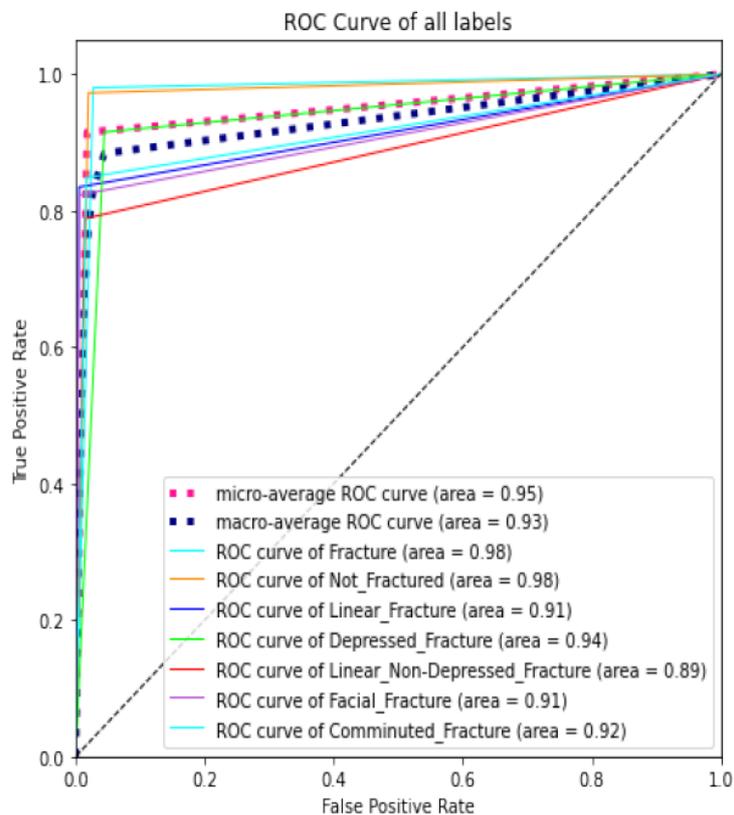

**Fig. 12:** ROC curves of the algorithm on the dataset

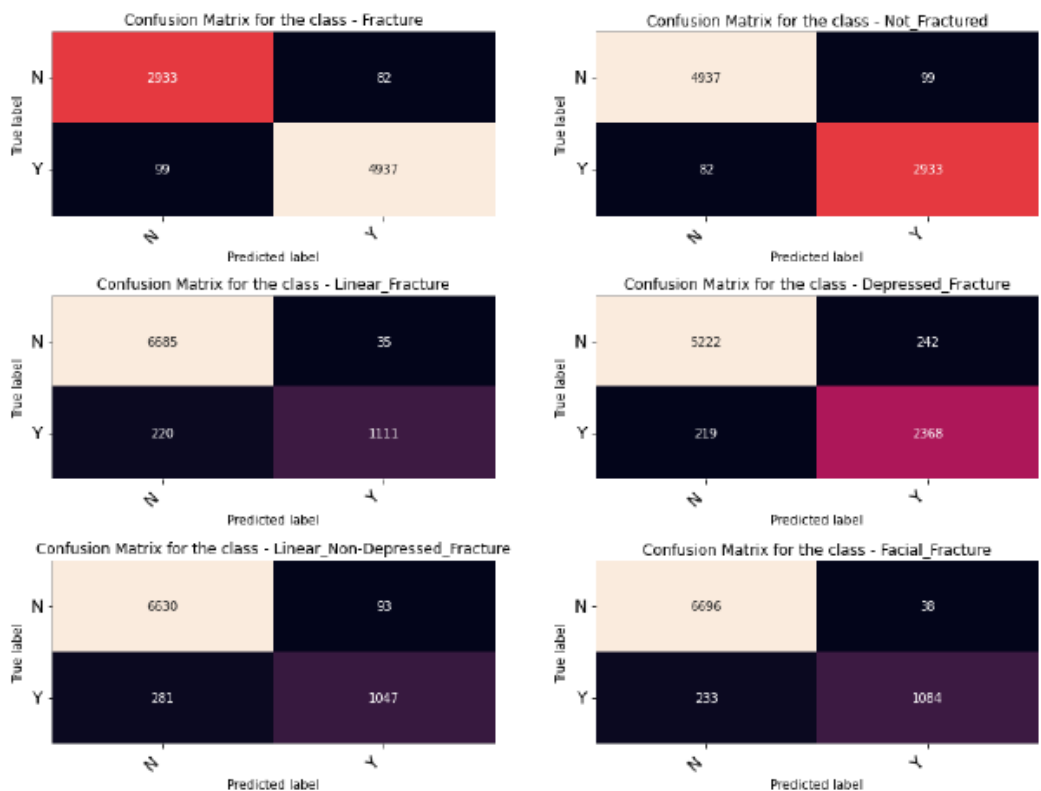

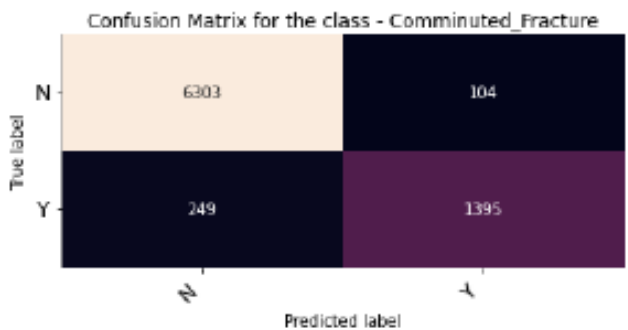

**Fig. 13:** Confusion matrix obtained for seven-class multi-label classification

The Area Under the Curve (AUC) of SkullNetV1 is between 0.89 and 0.98 (Fig. 12), which is quite promising. A summary of prediction results from the testing dataset by each class has been shown with Confusion Matrix in Fig. 13.

## Discussion

Radiology reports were used as ground labels in this study to increase classification performance over CT scan pictures alone and deep learning architectures with supervised learning were proposed for incorporating auxiliary data during training. The proposed method was successful in classifying a head CT scan image based on the type of skull fracture. A good deal of artificial intelligence-based research has been done to categorize a head CT picture as fractured or normal, but none have been done to further classify that image based on fracture type until this study. To clarify this, Table 4 shows a comparison with existing published reference models vs. our proposed work. Despite having fewer image data, our model performed really well. Our SkullNetV1 took only around 10 min to be trained, whereas other models took over 35 min, which indicates our model's time complexity is better than other mentioned models. When a larger training set is available, a larger model performs better. As a result, the number of layers we should use is determined by our data, the size of the dataset and the level of detail in our data. However, this does not imply that the more layers we have, the better. He *et al*. (2016) conducted a fascinating experiment in the original Res Net paper, they found that stacking more and more layers harmed the model's performance. By adding more layers, more parameters are introduced and the dataset may be too small to efficiently train these parameters. If the objective is simple and the data is limited, training it on a large complicated network will cause it to learn too much, resulting in overfitting. ResNet50 is a deeply layered CNN with 48 Conv layers.

Alex Net, on the other hand, is not a deeply layered model. In our experiment, we implemented the Alex Net model proposed in the original paper by Krizhevsky *et al*. (2012), which is made up of 5 Conv layers starting with an 11x11 kernel. A 121-pixel square receptive field can be achieved with an 11x11 kernel. A 5x5 kernel size, on the other hand, has a receptive field of 25 pixels in a square. This means that the kernel size of 11 will be able to incorporate more information during each dot product. However, the larger the kernel size does not necessarily mean the better it is, because, in our dataset, too many parameters lead to overfitting. This is what's going on with Res Net 50, Alex Net and other well-known models we've implemented. While these models perform well in large datasets and are extremely well trained in the "ImageNet" dataset, they struggle in small datasets. A small, imbalanced multi-label medical dataset has been used in our study. Due to the fact that "ImageNet" does not contain medical images, these larger models performed poorly in our dataset, prompting us to create our own baseline CNN with only 8 convolutional layers, which extracts features from our small dataset more effectively than notable transfer learning-based models, as shown in Fig. 9.

However, regardless of experimenting by tuning several hyperparameters of our baseline CNN, it still shows minor overfitting. This is most likely because our multi-label medical dataset is imbalanced and has a relatively low amount of images than a CNN would require. To overcome this, we developed our proposed model by substituting the dense layers of our baseline CNN with ML-KNN for the classification task. Our proposed model outperformed ResNet50, Alex Net and other prominent models because of this lazy learning approach combined with our baseline CNN. The extracted features of images by CNN were fed into a lazy learning-based classifier to classify the new data points based on the similarity measure of the earlier stored data points, which is a good strategy when the dataset is small. When a new test image comes it checks its similarity with earlier stored data. The findings of this study show that our architecture might increase medical image classification performance with high accuracy and F1 score.

Although our small dataset is accurate, there are fewer images of Linear Non-Depressed Fracture. Hence our baseline CNN combined with ML-KNN did not perform very well on this fracture type compared to other fractures, demonstrating that even with auxiliary data, training with a small dataset is difficult. As our proposed model consists of our baseline CNN which uses ML-KNN, the large memory requirement to store the training dataset is one of the drawbacks. The cost of calculating the Euclidian distance by our proposed model between the new point and each existing point is high which could degrade the performance of the algorithm if the dataset gets very larger. Understanding how a DNN produces predictions is a hot topic in medical research. It could persuade clinicians that the results are accurate, even if the model used an inaccurate area of the image instead of the genuine lesion site to generate the answer. CT scan images and radiology reports were used for training and CT scan images were used for testing in this study. The fracture types predicted from images by the SkullNetV1 were then visualized (Fig. 14 and 15).

Our model correctly predicted the fracture type in Fig. 14 and 15 shows how our model failed to identify the correct fracture type.

However, using radiology data as supplementary information to classify different skull fractures has never been done before. We started by classifying skull fractures and comparing them to other state-of-the-art CNN models. Although the SkullNetV1 performed well, the current study included some limitations. We had to exclude Basilar fracture, the most severe and rare type of skull fracture because CT scan images of this fracture were unavailable. The dataset was highly imbalanced. The GAN and ML-SMOTE algorithm could solve this problem. The deep Generative Adversarial Network (GAN) is used for data augmentation to combat data shortages and overfitting. It's a method of training an image generator model with an image discriminator model using large, unlabeled datasets. ML-SMOTE is one of the most well-known resampling algorithms which is used to create synthetic samples combining the features of samples from the minority classes with interpolation techniques. More data from Linear Non-Depressed Fracture could help us overcome our relatively poor performance in the detection of this fracture. As the experimental setup required a lot of computational power, different optimization strategies could be adopted to reduce the needed computational power.

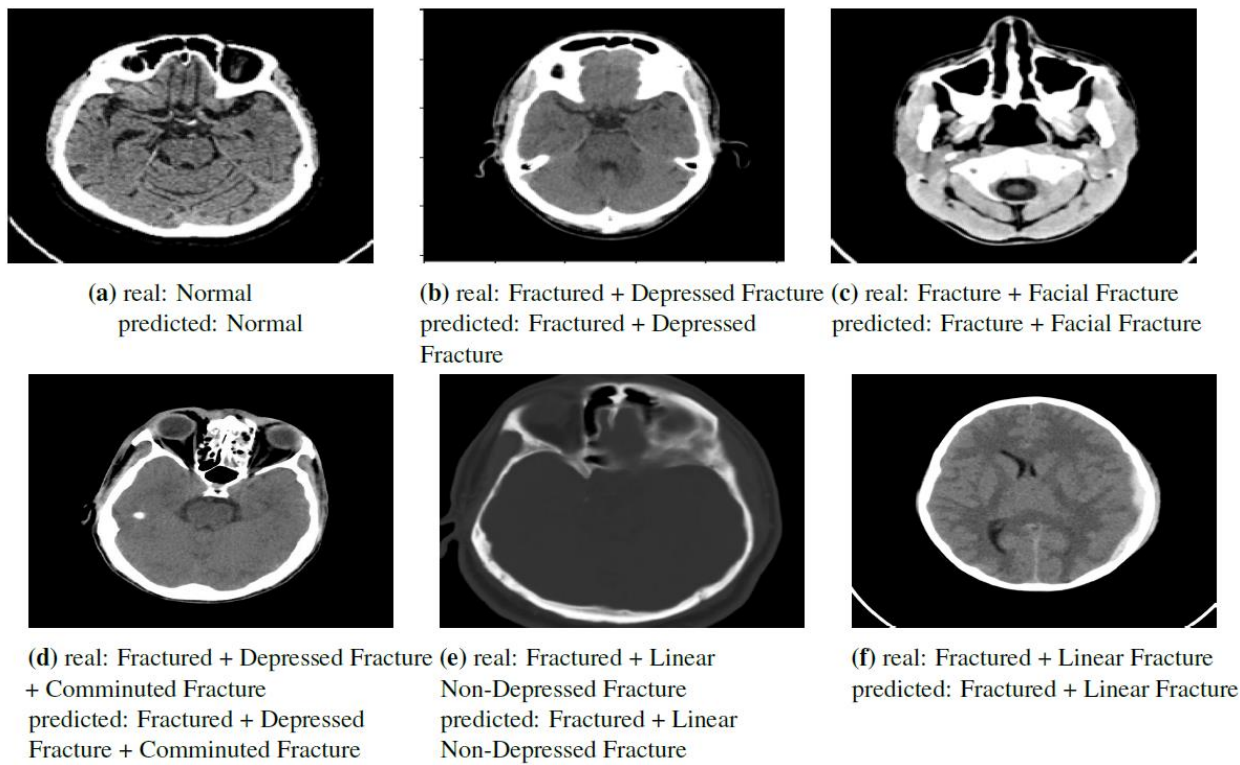

**Fig. 14:** Correct predictions

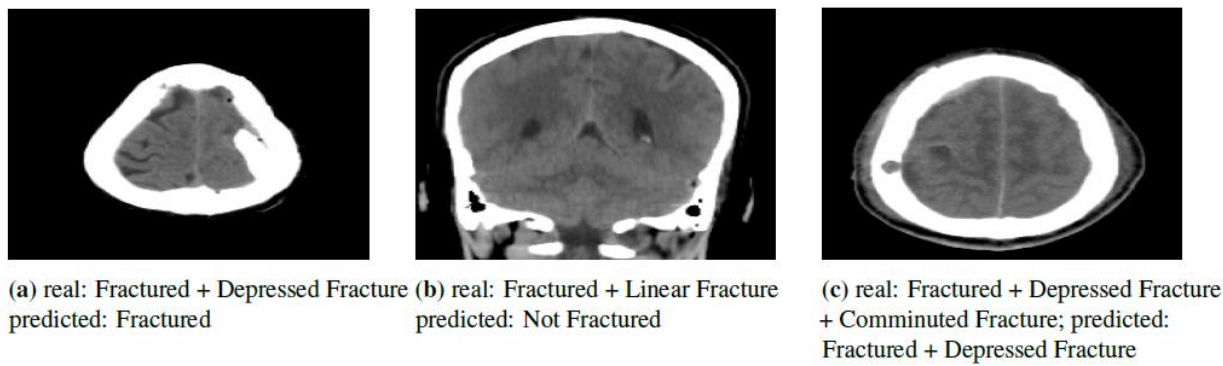

**Fig. 15:** Wrong predictions

## Conclusion

We developed a novel deep learning algorithm to automatically identify and classify skull fractures using head CT images. The proposed method for autonomously extracting relevant features from skull CT scans is rapid and reliable and training those features with a supervised learning algorithm is identical to the clinician's decision-making process. Compared to earlier studies on only skull fracture detection, our goal was to categorize skull fractures from images. Our model classifies skull fracture images trained with limited data. Advantages of our approach include simplicity and effective use of limited training data. For this seven-class multi-labeled classification, our recommended model has a subset accuracy of 88 percent, an F1 score of 93 percent, an AUC of 0.89 to 0.98, a Hamming score of 92 percent and a Hamming loss of 0.04. We hope that by applying our method to head CT images, we will automate the triage process of head CT scans. Our approach could help radiologists work more efficiently. This technique proved to have a promising future in applying medical image diagnosis in providing second opinions to doctors or radiologists. One of our significant areas of concentration for future study is the continuous improvement of the algorithm. Several enhancements might be made, such as the addition of GAN and the ML-SMOTE algorithm to balance the multi-label dataset.

## Acknowledgement

The authors are grateful to Syed Md. Shakawath Hossain, Medical Technologist, CT Scan department from Medinova Medical Services Ltd.; Md Sad Uddin Sadik, assistant manager; MD Ismail Hossain, Medical Technologist, CT Scan department from Ibn Sina Hospital Sylhet Limited for their support. They provided the CT brain images used in this study with respective authority permission. Dr. Tahmina Sumi from Z. H. Sikder Women's Medical College, Professor Dr. Ashikur Rahman Majumder, Head of the Department, Department of Radiology and Imaging from Sylhet MAG Osmani Medical College and Dr. Sajal Chandra Das from Ibn Sina Hospital Sylhet Limited, provided further annotations of CT scan images along with the respective radiologists. We also thank Sakib Alam Snigdha for lending us his computer in our initial computation.

## Author Contribution

**Md Moniruzzaman Emon:** Initiated, designed and executed the research. Collected the dataset, analyzed the results and improved the algorithm. Wrote the manuscript.

**Tareque Rahman Ornob:** Designed and executed the research. Structured the dataset to use it in the coding. Developed the algorithms, analyzed the results and made the necessary change in the algorithm. Helped in writing the manuscript.

**Moqsadur Rahman:** Supervised the research and gave required permission in data collection and guided the development of algorithms. Provided all the required equipment for the experimental setup.

## Ethics

This article is original and contains unpublished material. The corresponding author confirms that all of the other authors have read and approved the manuscript and no ethical issues are involved.